\documentclass{elsart3p}

\usepackage{graphicx}

\usepackage{amssymb}

\begin{document}

\begin{frontmatter}



\title{Superconductor strip in a closed magnetic environment: \\
exact analytic representation of the critical state}


\author{Y.A.~Genenko\corauthref{cor1}},
\corauth[cor1]{Corresponding author. }
\ead{yugenen@tgm.tu-darmstadt.de}
\author{H. Rauh}
\address{Institut f\"{u}r Materialwissenschaft, Technische Universit\"{a}t Darmstadt, 64287 Darmstadt, Germany}

\begin{abstract}
An exact analytic representation of the critical state of a current-carrying type-II superconductor strip located inside a 
cylindrical 
magnetic cavity of high permeability is derived. The obtained results show that, when the cavity radius is small,  
penetration of magnetic flux fronts is strongly reduced as compared to the situation in an isolated strip.  
From our generic representation it is possible to establish current profiles in closed cavities of various other 
geometries too by means of conformal mapping of the basic configuration addressed.          

\end{abstract}

\end{frontmatter}

Relatively high AC losses in superconductor cables and strips present a substantial problem for the implementation of superconductors 
in high-frequency and low-frequency applications. Recently, a suggestion for improving the current-carrying capability of 
superconductor 
strips \cite{Genenko1,Genenko2,GRS} and for reducing AC losses in superconductor cables \cite{Campbell1} based on the idea of magnetic
shielding of applied fields as well as current self-induced fields was put forth. AC losses in superconductor strips caused by the 
latter type of fields are anticipated to greatly decrease when the strips are exposed to suitably designed magnetic environments 
\cite{Genenko2,GRS}.
Exact analytic representations of sheet current distributions in superconductor strips located between two high-permeability magnets 
occupying infinite half-spaces were derived before \cite{Genenko1,Genenko2}; these configurations allowed to find the respective 
current distributions for various other topologically {\it open} shielding geometries by application of the method of conformal 
mapping. Utilization of the latter tool for analyzing sheet current distributions and AC losses in the presence of topologically 
{\it closed} magnetic environments of practical 
interest requires corresponding reference results. An establishment of such results is the focus of the present communication.\

We consider an infinitely extended type-II superconductor strip of width $2w$ located inside a cylindrical cavity of radius $a$ 
in an infinitely extended soft magnet of relative permeability $\mu$, the symmetry axis of this configuration coinciding with 
the $z$-axis of a cartesian coordinate system $x, y, z$. Assuming the thickness of the strip to be small 
compared to its width, variations of the current over the thickness of the strip may be ignored and, for mathematical convenience, 
the state of the strip characterized by the sheet current $J$ alone.

When magnetic flux penetrates the superconductor strip in the critical state, the distribution of the sheet current is controlled 
by the pinning of magnetic vortices. In conformity with Bean's hypothesis \cite{Bean}, the sheet current adopts 
its critical value $J_c$  throughout the flux-penetrated regions of the strip, whereas the magnetic field component normal to the 
strip vanishes in the flux-free regions of the strip. Proceeding in the spirit of previous work \cite{Norris}, a distribution of the 
sheet current prevails with magnetic flux penetrated from the edges of the strip, but with the central zone $-b<x<b$ of half-width   
$b<w$  left flux free. In this zone, the distribution of the sheet current is governed by the integral equation \cite{Genenko2}

\begin{equation}
\label{maineq}
\int_{-w}^{w}dx' J(x') \left( \frac{1}{x-x'} + \frac{q}{x-a^2/x'} \right) = 0
\end{equation}

\noindent together with the requirement that the total
transport current equals $I$. 
%
%
\noindent Here, $q=(\mu-1)/(\mu+1)$ means the strength of the image current induced by the magnetic cavity. 

In the limit $\mu\gg 1$, {\it i.e.} for $q\rightarrow 1$, Eq. (\ref{maineq}) has 
the exact analytic solution 

\begin{equation}
\label{solution}
J(x)=J_c \left(\frac{a^2+b^2}{\pi s^2(x)}\right)  s'(x) \phi(s(x)), 
\end{equation}

\noindent where

\begin{eqnarray*}
\phi(s)=\frac{c}{\sqrt{c^2-s^2}}\arctan{\sqrt{\frac{(h^2-b^2)(c^2-s^2)}{(b^2-s^2)(c^2-h^2)}}}\phantom{123456}\\
-\,\arctan{\sqrt{\frac{h^2-b^2}{b^2-s^2}}} +\frac{\sqrt{b^2-s^2}}{b}\arctan{\sqrt{\frac{b^2w^2-b^4}{a^4-b^2w^2}}}\\
\times \left[ \left(K\left(\frac{b}{c}\right)-\left.\Pi\left(\frac{b^2}{s^2},\frac{b}{c}\right)\right)\right/K\left(\frac{b}{c}\right) 
\right]\phantom{1234567890}
\end{eqnarray*}

\noindent with $s(x)=x(a^2+b^2)/(a^2+x^2),\,c=(a^2+b^2)/2a$ and $h=w(a^2+b^2)/(a^2+w^2)$. Herein, $K$ and $\Pi$ denote complete
elliptic integrals of the first and, respectively, third kind.

Sheet current profiles obtained from Eq. (\ref{solution}) for a range of the geometrical parameters involved, with a fixed 
value of $b$, are shown in Fig. 1. This exhibits a flattening of the current profiles together with an increase in the magnitude 
of the total current up to saturation, when the radius of the magnetic cavity is reduced, precisely as in the case of topologically 
open magnetic cavities \cite{Genenko1,Genenko2}. 

\begin{figure}[h]
\begin{center}
\includegraphics[width=7.5cm]{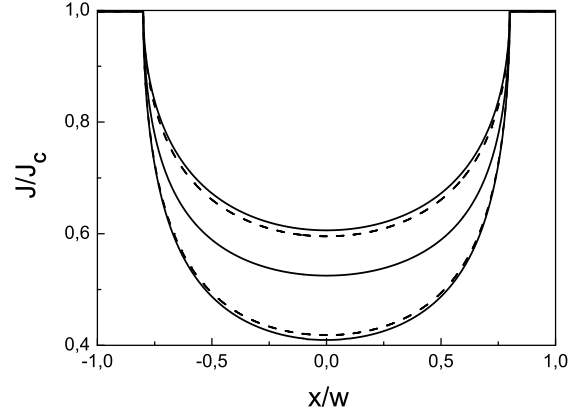}
\caption{ Distribution of the sheet current over the flux-free zone of the partly flux-filled strip delineated by
$b/w=0.8$, with $a/w=1.001, 1.01, 1.1, 2$ and infinity (from the upper curve down).}
\label{finequid0caca}
\end{center}
\end{figure}

The half-width of the flux-free zone is controlled by the total transport current in the strip and by the geometry of the 
magnetic environment. An implicit equation for $b$ in the chosen geometry may be found by integrating the sheet current over the
width of the strip using Eq. (\ref{solution}) which yields

\begin{equation}
\label{total}
I=\frac{J_c (\pi/b) (a^2+b^2)}{ K^2\left(2ab/(a^2+b^2)\right)} \arctan{\sqrt{\frac{b^2(w^2-b^2)}{(a^4-w^2b^2)}}}  .
\end{equation}

The cylindrical magnetic cavity entails a reduction of the depth of penetration of magnetic flux into the strip,  
$\Delta(I)=w-b$, as compared to the depth in the situation without a magnetic environment, $\Delta_0(I)=w(1-\sqrt{1-(I/I_c)^2})$,  
where $I_c=2wJ_c$ \cite{Brandt}. For weak flux penetration, when $\Delta \ll w$ and hence $I\ll I_c$,
the explicit approximate result

\begin{equation}
\label{penetration}
\Delta(I)\simeq\frac{2w}{\pi^2} \left(\frac{a^2-w^2}{a^2+w^2}\right) K^2\left(\frac{2aw}{a^2+w^2}\right) \left(\frac{I}{I_c}\right)^2  
\end{equation}

\noindent is seen to hold. Thus, $\Delta$ strongly decreases with respect to $\Delta_0\simeq (w/2) (I/I_c)^2$ as  
$a\rightarrow w$. This also means a reduction of AC losses to the same extent which typically scale with $\Delta ^2$ 
\cite{Norris,Brandt}. These losses may be further curtailed by optimization of the shape of the magnetic cavity 
using, in the limit $\mu \gg 1$, the method of conformal mapping of the basic cylindrical configuration addressed above.

\end{document}